\documentclass{aa}  

\usepackage{xcolor}

\usepackage{graphicx}

\usepackage{txfonts}
\usepackage{hyperref}

\begin{document}

   \title{Expanding on the Fundamental Metallicity Relation in Dwarf Galaxies with MUSE\thanks{Based on observation collected at the ESO Paranal La Silla Observatory, Chile, Prog. ID 0108.B-0904, 0104.D-0503, 0100.B-0116, and 095.B-0.532.}}
   
   \titlerunning{The Fundamental Metallicity Relation in Dwarf Galaxies}
   \subtitle{}
      \author{Teodora-Elena Bulichi
          \inst{1}\fnmsep\inst{2}
          \and
          Katja Fahrion
          \inst{1}
          \and
          Fran\c{c}ois Mernier
          \inst{1,3}
          \and
          Michael Hilker
          \inst{4}
          \and
          Ryan Leaman
          \inst{5}
          \and
          Mariya Lyubenova
          \inst{4}
          \and
          Oliver M\"uller
          \inst{6}
          \and
          Nadine Neumayer
          \inst{7}
          \and
          Ignacio Martin Navarro
          \inst{8,9}
          \and
          Francesca Pinna
          \inst{7}
          \and
          Marina Rejkuba
          \inst{4}
          \and
          Laura Scholz-Diaz
          \inst{8,9}
          \and
          Glenn van de Ven
          \inst{5}
    }
   
   \institute{European Space Agency (ESA), European Space Exploration and Research Centre (ESTEC), Keplerlaan 1, 2201 AZ Noordwijk, the Netherlands. \and Institute for Astronomy, University of Edinburgh, Royal Observatory, Blackford Hill, Edinburgh
EH9 3HJ, UK. \and SRON Netherlands Institute for Space Research, Niels Bohrweg 4, 2333 CA Leiden, The Netherlands.\and European Southern Observatory, Karl-Schwarzschild-Strasse 2, 85748 Garching bei M\"{u}nchen, Germany \and Department of Astrophysics, University of Vienna, T\"urkenschanzstrasse 17, 1180 Wien, Austria. \and Institute of Physics, Laboratory of Astrophysics, Ecole Polytechnique F\'{e}d\'{e}rale de Lausanne (EPFL), 1290 Sauverny, Switzerland.\and Max-Planck Institute for Astronomy, K\"{o}nigstuhl 17, 69117 Heidelberg, Germany.\and  Instituto de Astrofisica de Canarias, E-38205 La Laguna, Tenerife, Spain. \and Departamento de Astrofısica, Universidad de La Laguna, E-38205 La Laguna, Tenerife, Spain.}

   \date{\today}

 
  \abstract
   {The mass-metallicity relation (MZR) represents one of the most important scaling relations in the context of galaxy evolution, comprising a positive correlation between stellar mass and metallicity (Z). The fundamental metallicity relation (FMR) introduces a new parameter, the star formation rate (SFR), in the dependence. While several studies found that Z is anti-correlated with the SFR at fixed mass, the validity of this statement has been questioned extensively and no widely-accepted consensus has been reached yet. With this work, we investigate the FMR in nine nearby, spatially-resolved, dwarf galaxies, using gas diagnostics on integral-field spectroscopic data of the Multi Unit Spectroscopic Explorer (MUSE), pushing such investigations to lower galaxy masses and higher resolutions. 
   We find that
   both the MZR and FMR exhibit different behaviours within different star forming regions of the galaxies. We find that the SFR surface density - metallicity anti-correlation is tighter in the low-mass galaxies of our sample. 
   For all the galaxies considered, we find a SFR surface density - stellar mass surface density correlation. We propose that the main reason behind these findings is connected to the accretion mechanisms of the gas fuelling star formation -- low-mass, metal-poor galaxies accrete pristine gas from the intergalactic medium, while in more massive and metal-enriched systems the gas responsible for star formation is recycled from previous star forming episodes.}


  \maketitle


\section{Introduction}
\label{section:introduction}

Metallicity is a key factor describing galaxy properties, as it contains important information about the physical processes involved in their formation and evolution. Metallicity is influenced by infall of metal-poor gas, star formation and outflow of enriched material processed in stars. These cyclic processes are all crucial drivers of the overall galactic evolution. Consequently, a great amount of effort has been put into the analysis of scaling relations connecting the metal-content of galaxies to their global properties, such as the mass-metallicity relation (MZR, see \citealt{review_paper} for a detailed review). The pioneer papers in this area, such as \cite{mcclure_vandenBergh,sandage1972,mould_et_al_1983} and \cite{buananno_et_al_1985} investigated the stellar metallicity and mass correlation by color magnitude diagrams and stellar spectroscopy. Similarly, \cite{lequeux_et_al_1979} and \cite{skillman_1992} studied the connection between stellar mass and gas-phase metallicity. These works clearly showed a correlation between stellar mass and metallicity (both stellar and gas-phase) that implies that more massive galaxies are more metal-enriched. This has been confirmed observationally in local quiescent and star-forming galaxies, with the aid of the extensive Sloan Digital Sky Surveys (SDSS) spectroscopic database and other large spectroscopic galaxy samples (e.g. \citealt{trager_et_al_2000b,kuntschner_et_al_2001,gallazzi_et_al_2005,thomas_et_al_2005,gallazzi_et_al_2008,panter_et_al_2008,graves_et_al_2009,thomas_et_al_2010a,harrison_et_al_2011,Petropoulou_et_al_2011,kirby_et_al_2013,conroy_et_al_2014,Gonzales_delago_et_al_2014,fitzpatrick_graves_2015,Sybilska_et_al_2017,zahid_et_al_2017,lian_et_al_2018a,zhang_et_al_2018b}), as well as in semi-analytical models and hydrodynamical simulations (e.g. \citealt{de_lucia_et_al_2004,croton_et_al_2006,finlator_dave_2008,oppenheimer_dave_2008,dutton_et_al_2011,dave_et_al_2011,somerville_et_al_2012,pipino_et_al_2014,torrey_et_al_2014,zahid_et_al_2014,feldmann_2015,lu_et_al_2015,kacprzak_et_al_2016,christensen_et_al_2016},\mbox{\citealt{hirschmann_et_al_2016}},\citealt{rodriguez-puebla_et_al_2016,torrey_et_al_2018, simba})

Nowadays, the general trend of the MZR has been reported for a broad range of galaxy masses. However, significantly more focus has been put on massive galaxies ($\gtrsim$ 10$^9$ M$_{\text{\sun}}$), because these galaxies are brighter and hence more accessible observationally. Despite the extensive research in this area, there is ongoing investigation of the physical processes behind the MZR and its shape. The following five hypotheses have been proposed to explain the MZR: i) the loss of enriched gas by galactic outflows from stellar feedback, which is expected to remove a considerable fraction of metal-enriched gas, especially from low-mass systems, towards the circumgalactic medium (CGM) and intergalactic medium (IGM) \citep{tremonti_et_al_2004,kirby2011,tumlinson_et_al_2011}; ii) the accretion of less-enriched gas by inflows \citep{finlator_dave_2008,rodrigues_et_al_2012,lagos_et_al_2016b}; iii) variations of the initial mass function (IMF) with galaxy
mass, affecting therefore the average rate of metal enrichment \citep{trager_et_al_2000b, koppen_et_al_2007, molla_et_al_2015,vincenzo_et_al_2016,lian_et_al_2018}; iv) selective star-formation efficiency
or downsizing, suggesting that high-mass galaxies are favoured to form (over low-mass ones) at higher redshift, hence enabling higher-mass galaxies to have formed more metal-enriched gas \citep{brooks_et_al_2007, ellison_et_al_2008, maiolino_et_al_2008, calura_and_2009, zahid_et_al_2011} and v) higher metallicity of the accreted gas for higher mass galaxies -- as a result of the recycling from previous episodes of star formation \citep{brook_et_al_2014,ma_et_al_2016}. 

In order to explore the underlying phenomena that result in the observed MZR, additional factors have been investigated once the instrumentation became precise enough. For example, \cite{tremonti_et_al_2004} reported that the MZR presents secondary correlations with galaxy colour, ellipticity and central mass density; \cite{hoopes_et_al_2007} stated that the galaxy size influences the shape of the MZR and \cite{ellison_et_al_2008} proposed the idea that the star formation rate (SFR) is connected with the metallicity at a given mass. \cite{mannucci_et_al_2010} introduced the concept of the fundamental metallicity relation (FMR) that describes the relation between mass, metallicity and SFR, with two notable effects: i) at fixed mass, galaxies with higher SFRs have lower metallicities; and ii) for low-mass galaxies, the metallicity dependence on the SFR is stronger. 

The FMR appeared to be in contradiction with the star formation main sequence (SFMS; e.g. \citealt{noeske_et_al_2007,daddi_et_al_2007, Dave_2008, johnston_et_al_2015}), which finds that the SFR is proportional to the stellar mass of the galaxy, which in turn is proportional to the metallicity. Within the possible explanations for the metallicity-SFR anti-correlation, \cite{mannucci_et_al_2010} proposed that the infall of metal-poor gas fuels star formation. In this scenario, at a given mass, a high SFR would imply a high accretion rate of metal-poor gas that lowers the overall metallicity. The exact mechanisms behind this anti-correlation are still not universally accepted. However, great efforts have been undertaken in testing this relation both observationally (e.g. \citealt{cresci_et_al_2010, hwang_et_al_2019}) and in numerical simulations (e.g. \citealt{sanchez_almeida_et_al_2017}), with contradicting results. For example, within the large integral field spectroscopy (IFS) surveys (CALIFA -- \citealt{sanchez_et_al_2012a}, SAMI -- \citealt{croom_et_al_2012}, MaNGA -- \citealt{bundy_et_al_2015}),  \cite{BarreraBallesteros_et_al_2017} and \cite{sanchez_et_al_2017} report no MZR-SFR dependence, while \cite{rosales_ortega_et_al_2012} found a SFR dependence in their analysis. 

Furthermore, IFS surveys enabled spatially resolved studies of star formation and metal-enrichment processes (e.g. \citealt{paper_fNacho, Kumari2019, birkin2023jwsts}). They allow us to understand whether the observed relations emerge from averaging the local contributions to the total mass, metallicity and SFR (see also \citealt{baker}). For example, \cite{Gonzales_delago_et_al_2014} and \cite{BarreraBallesteros_et_al_2017} found a clear correlation between metallicity and stellar mass surface density ($\Sigma_{*}$), investigating galaxies at the high-mass end (M $\gtrsim$ 10$^9$ M$_{\text{\sun}}$) from the CALIFA and MaNGA surveys, respectively.  However, the contribution of the star formation (often measured as a star formation rate density $\Sigma_{\mathrm{SFR}}$) remains debated.

The next generation of IFS instruments such as the Multi Unit Spectroscopic Explorer (MUSE) allows us to take the next step due to the superior sensitivity, wide field of view and spatial resolution.  MUSE enables the study of these scaling relations in unprecedented detail in nearby galaxies (e.g. \citealt{erroz-ferrer_et_al_2019,buzzo_et_al_2021,lara-lopez_et_al_2022}) and is capable of providing a broader perspective in the context of the galaxies at the low-mass end, which have not been explored extensively.

There are still open questions regarding the contribution of the SFR to metallicity scaling relations, but also the universality and origins of the MZR and the FMR. With this work, we make use of MUSE's excellent sensitivity to investigate the existence of a spatially resolved FMR in low-mass galaxies, analysing various different behaviours within the individual galaxies in our nine local dwarf galaxies sample. The paper is structured as follows: Section~\ref{section:galaxy_sample} presents the galaxy sample and an overview of the general properties of the galaxies selected for this study; Section~\ref{sect:rebinned} describes the process of rebinning our data to a suitable resolution to enable i) the use of the methods described in Sect.~\ref{section:fit_to_emission_lines}  and ii) relevant comparisons with previous literature results; Section~\ref{section:fit_to_emission_lines} introduces the methods to derive the quantities of interest (metallicity, surface star formation rate density and stellar mass density); Section~\ref{section:FMR} describes how the spectral fitting yielded to the results needed for the FMR analysis, while Sect.~\ref{section:conclusions} discusses our findings in comparison with previous literature studies and Sect.~\ref{conclusions_2.0} concludes the paper.

\begin{table*}[!t]
\caption{Galaxy properties}\vspace{-4mm}
\begin{center}
\begin{tabular}{c c c c c c c c c c} \hline\hline
Galaxy & Morphology & RA & DEC & D & E(B-V) & log(M$_\text{gal}$) &  z & Ref   \\
    &  &  & & (Mpc) & (mag) & log(M$_{\text{\sun}}$) &   &  \\ 
(1) & (2) & (3) & (4) & (5) & (6) & (7) & (8) & (9)  \\    \hline
ESO\,059-01& IB(s)m & 07:31:18.20 & $-$68:11:16.8 & 4.57 & 0.147 & 8.1  &  0.00177 & a, b, c, d\\
IC\,1959 & SB(s)m & 03:33:12.59 & $-$50:24:51.3 & 6.05 & 0.011 & 8.2  & 0.00213 & a, b, c, d  \\
NGC\,1487 & pec &03:55:46.10 & $-$42:22:05.0 & 12.1 & 0.010 & 8.95 & 0.00283 &  a, e, f \\
NGC\,1796 & SB(rs)c pec? &05:02:42.55 & $-$61:08:24.2 & 10.6 & 0.021 & 9.06  & 	 0.00338&  a, e, f \\
NGC\,4592 & SA(s)dm & 12:39:18.74 & $+$00:31:55.2 & 10.6 & 0.042 & 9.02  & 	 0.00357 & a, e, f \\
NGC\,853 & Sm pec? &02:11:41.19 & $-$09:18:21.6 & 21.0 & 0.023 & 6.96  & 0.00501 & a, e, f\\
UGC\,3755& 	Im & 07:13:51.60 & $+$10:31:19.0 & 4.99 & 0.078 & 7.75  & 0.00105&  a, e, f \\
UGC\,5889 & SAB(s)m & 10:47:22.30 & $+$14:04:10.0 & 6.89 & 0.031 & 8.20 &  0.00191 & a, e, f \\
UGC\,8041 & SB(s)d & 12:55:12.66 & $+$00:06:59.9  & 17.1 & 0.019 & 8.83  & 0.00442 & a, e, f \\
 \hline
\end{tabular}
\label{tab:galaxies}
\end{center}
\tablefoot{(1) galaxy name, (2), morphology (3), (4) right ascension and declination, (5) distance in Mpc, (6) foreground extinction, (7) galaxy stellar masses, (8) redshift from from the NASA/IPAC Extragalactic Database (NED) and (9) references: a - \cite{morphology} b - \cite{Georgiev2009a}, c - \cite{Georgiev2009b}, d - \cite{Yu2020},  f - \cite{Georgiev2014}, e - \cite{Georgiev2016}}
\end{table*}

\section{Galaxy Sample}
\label{section:galaxy_sample}
\begin{figure*}
    \centering
    \includegraphics[width=0.95\textwidth]{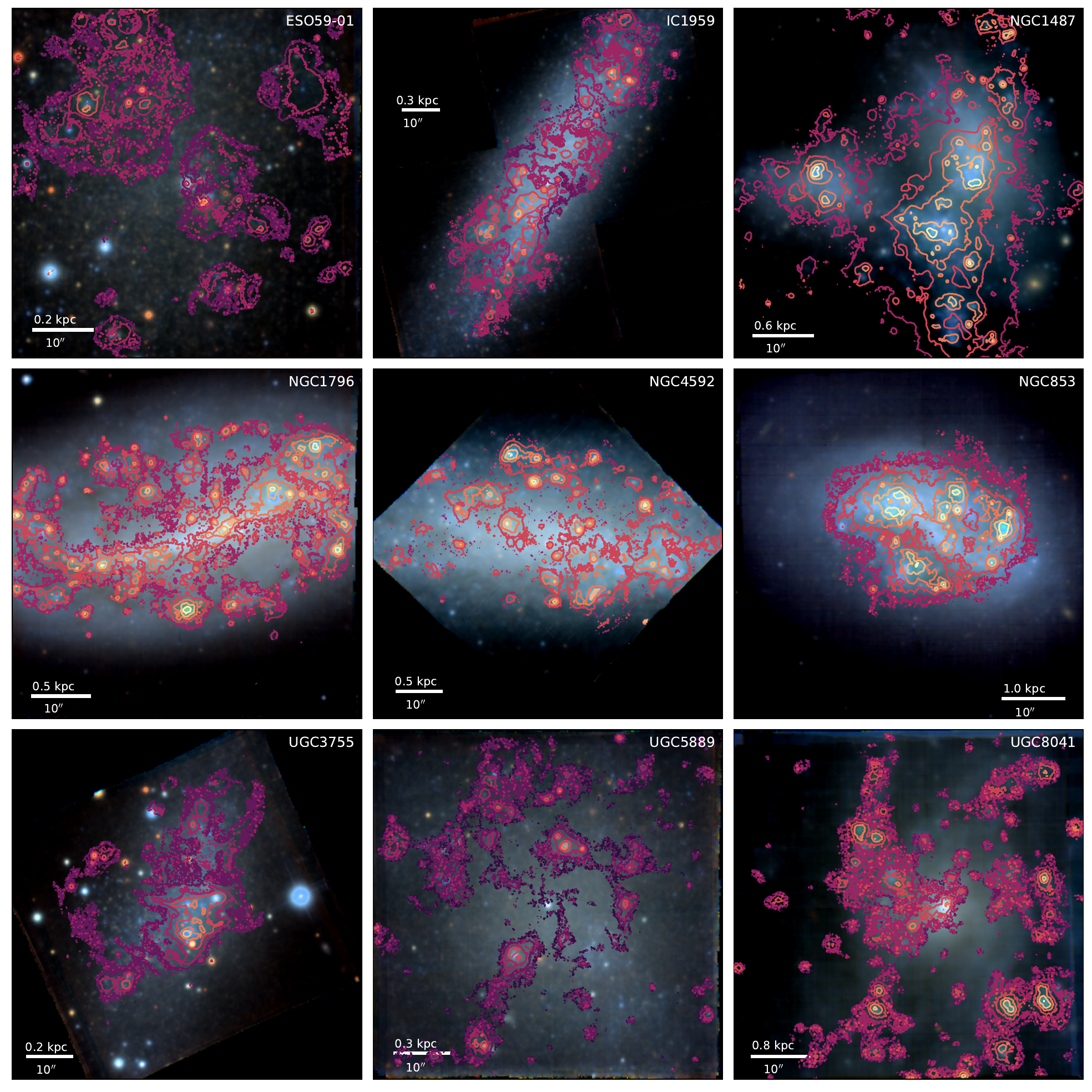}
    \caption{RGB images of the nine galaxies in our sample obtained from the MUSE cubes. The purple lines mark extinction-corrected H$\alpha$ flux contours as obtained from running DAP on the original data. The contours mark fluxes from 10$^{-18}$ to 10$^{-15}$ erg s$^{-1}$ in 0.5 dex intervals.}
    \label{fig:RGB}
\end{figure*}

In this study we used the same nine galaxies as in \cite{Fahrion2022} that presents a detailed analysis of the nuclear star clusters properties and host galaxy star formation histories. In the present work, we provide a complementary analysis of the gaseous components, with a specific focus on the FMR. All of the galaxies selected for this paper are nearby (distances $\leqslant$ 21 Mpc) late-type dwarf galaxies, observed with MUSE. For reference, we show the RGB images with H$\alpha$ contours, in Fig.~\ref{fig:RGB}. 

The MUSE integral field spectrograph (\citealt{bacon2010}) is mounted on the Very Large Telescope of the European Southern Observatory (ESO). MUSE data cover the optical wavelength range between 4700 and 9300 $\AA$ at a field of view (FoV) of 1 arcmin$^{2}$ and contain a total of 90000 spaxels, each with about 3800 spectral pixels sampled at 1.25 \AA~per pixel with a mean instrumental resolution of $\sim$ 2.5~\AA.  Our sample contains a total of nine galaxies: ESO\,59-01, IC\,1959, NGC\,1487, NGC\,1796, NGC\,853, UGC\,3755, UGC\,5889, NGC\,4592 and UGC\,8041. The first seven were observed in 2020 and 2021 as part of the MUSE programme 0108.B-0904 (PI: Fahrion) making use of adaptive optics-supported wide-field mode. The last two were retrieved from the ESO Science Archive. NGC\,4592 was originally observed as part of programme 095.B-0532 (PI: Carollo) in 2015
for the MUSE Atlas of Disks (MAD) survey \citep{erroz-ferrer_et_al_2019}, while UGC\,8041 was observed under programme
0104.D-0503 (PI: Anderson) in 2020. The nine galaxies display different morphologies and NGC\,1487 deserves a special note given that it is an ongoing merger (see e.g. \citealt{buzzo_et_al_2021} for a more detailed description of this galaxy). The galaxies span a mass range between $\sim$10$^7$-$10^9$ M$_\sun$. The general properties of the nine galaxies are presented in Table~\ref{tab:galaxies}.

For the seven galaxies part of the 0108.B-0904 programme (more details can be found in \citealt{Fahrion2022}), the data were reduced via the MUSE data reduction pipeline version 2.8.5 \citep{Weilbacher_et_al_2020}, within the esorex framework version 3.13.5. Except for NGC 1487 and IC 1959, all galaxies were covered with one pointing with exposure times ranging between one and four hours. To cover the disk of IC 1959, two pointings were used and by combining the 2021 MUSE data with archival data from 2018 (programme 0100.B-0116, PI: Carollo), we obtained a three pointing mosaic for NGC 1487.

\section{Rebinning the data to 100 pc}
\label{sect:rebinned}
Due to the small distances to our sample galaxies ranging between 4 and 20 Mpc, individual MUSE spaxels of 0.2\arcsec $\times$ 0.2\arcsec per pixel correspond to spatial scales between 4 and 20 pc. These scales are smaller than the typical HII regions for which the methods to obtain gas phase metallicities and star formation rates have been established (e.g. \citealt{sanchez_review, sanchez2021rmxaa}). We note that some previous literature results have argued that the resolved SFMS and Schmidt-Kennicutt (SK) law emerge from the self-regulation of the star formation process, finding that these relations break at scales lower than the typical size of the largest molecular clouds ($\sim$ 500 pc, e.g. \citealt{kruijssen, sanchez2021mnras}).

\begin{figure*}[h!]
    \centering
    \includegraphics[width=1\textwidth]{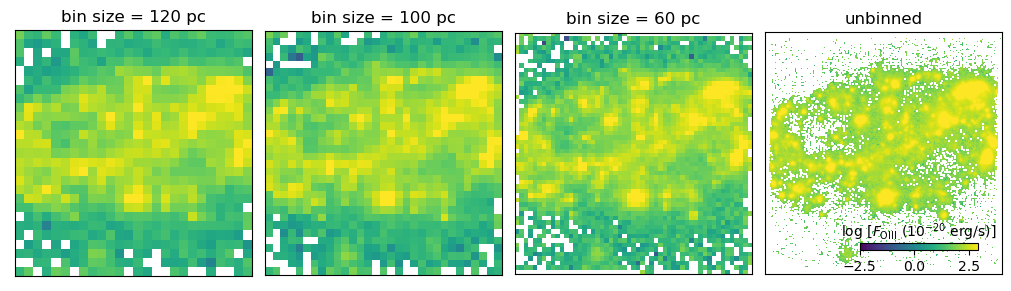}
    \centering
    \caption{The [OIII] flux maps for NGC\,1796, using rebinned (degraded) data -- first three panels from the left, as follows: bin size = 120 pc, 100 pc, 60 pc. The last panel shows the same [OIII] flux using the unbinned data. Each panel shows the 1 $\times$ 1\arcmin\,field of view of MUSE.}
    \label{fig:comp_maps}
\end{figure*}

On the other hand, \cite{Pessa2021} have shown that the resolved SFMS, SK and the resolved molecular gas main sequence are recovered at scales as low as 100 pc, using galaxies from Physics at High Angular Resolution in Nearby GalaxieS (PHANGS) survey. While the scatter increases at smaller scales, the proximity of our galaxy sample does not allow us to rebin to scales of ~500 pc or larger without removing all spatial resolution provided by MUSE. As such, given our goal to study the resolved FMR in these dwarf galaxies, pushing earlier studies to lower masses and higher resolutions, we rebinned our MUSE data to a common size of 100 pc per bin using \textsc{MPDAF} \citep{mpdaf}. Figure~\ref{fig:comp_maps} shows this rebinning on the example of NGC\,1796 which is at an intermediate distance of 10 Mpc. This figure shows the [OIII] flux map obtained as described above using different bin sizes. For the subsequent analysis we only use the rebinned data to a bin size of 100 pc. We note that rebinning only the derived maps of line fluxes leads to the same results.


\section{Spectral Fitting}
\label{section:fit_to_emission_lines}

We describe our methods to extract the relevant quantities such as gas-phase metallicity, star formation rate and stellar mass surface density in the following. We applied these methods to the rebinned data as described in Sect. \ref{sect:rebinned}.

\subsection{Emission line analysis with DAP}
We performed the emission lines analysis with the Data Analysis Pipeline (\textsc{DAP})\footnote{\url{https://gitlab.com/francbelf/ifu-pipeline}}, developed for the MUSE's Physics at High Angular Resolution in Nearby Galaxies Survey (PHANGS, for a detailed description of \textsc{DAP} see \citealt{Emsellem_et_al_2022}). \textsc{DAP} is a publicly available code for production of high-level data products from IFS observations of stellar absorption and ionized gas emission lines in galaxies. \textsc{DAP} has a modular framework that adapts different steps for the data analysis, including \textsc{vorbin} \citep{cappellari_copin_2003} for binning the data with the established Voronoi binning scheme, and \textsc{pPXF} \citep{cappellari_emsellem_2004, cappellari2017}, a full spectrum fitting method to fit the stellar absorption and gas emission lines. 

In the first step, \textsc{DAP} corrects the full datacube for foreground extinction along the line-of-sight using the \cite{cardelli_et_al_1989} extinction law. We used the E(B-V) values listed in Table \ref{tab:galaxies}. Then the data are binned to a target S/N of 90 to ensure a continuous S/N distribution across the FoV that is sufficient for an accurate measurement of the stellar population properties. The binned spectra are then fitted with \textsc{pPXF} to obtain the line-of-sight velocity distribution parameters (velocity, velocity dispersion) of the stellar component by cross-correlating with user-supplied stellar population models. In this step, also binned values for the emission line kinematics and fluxes can be obtained. However, to acquire a more detailed view, we ran the emission line analysis on a spaxel-by-spaxel basis. Here, \textsc{DAP} fits the spectra and in particular their emission lines of the individual spaxels while keeping the stellar velocity fixed to the value of each respective Voronoi bin. As output, \textsc{DAP} provides maps of binned stellar kinematics as well as fluxes and velocities of a range of different emission lines. 


\subsection{Stellar Mass Surface Density}
We used \textsc{DAP}'s capabilities to perform stellar population analysis to further obtain maps of the stellar mass surface density. To do so, \textsc{pPXF} within \textsc{DAP} fits E-MILES simple stellar population models \citep{vazdekis_et_al_2016} to the stellar spectra. We used the models provided within \textsc{DAP} assuming a \cite{chabrier_2003} IMF, BaSTI isochrones \citep{pietrinferni_et_al_2004}, with 13 ages (0.03 -- 13.5 Gyr) and six stellar metallicities ([M/H] = [$-$1.49, $-$0.96, $-$0.35, 0.06, 0.26, 0.40]). As the E-MILES models are normalised to 1 M$_\sun$, the best-fitting combination of models not only gives the mean age and stellar metallicity of each bin, but can also be used to infer the  mass surface density. In the stellar population fits, extinction in the stellar absorption spectrum is taken into account by using a two step approach: in the first iteration, the spectrum is only fitted for the extinction. This extinction is then kept fixed and the spectrum is fitted again for the stellar population parameters.

For the purpose of this study, we computed the stellar mass surface density, using Voronoi-bins with a signal-to-noise (S/N) threshold of 90 in order to get reliable stellar population properties. With this threshold, we found that the uncertainties associated with the stellar mass surface densities are $< 10\%$. These uncertainties refer to random uncertainties that are obtained by fitting the spectrum in each bin in a Monte-Carlo fashion 10 times (see \citealt{Emsellem_et_al_2022} for details). We also explored lower thresholds and obtained mean uncertainties around 35$\%$  when implementing a S/N threshold of 10, for example. A higher threshold, even though it would lower the uncertainties even further, would restrict considerably the number of data points. 
A more detailed stellar population analysis will be done in future work. Here, we only make use of the stellar surface mass density.



\subsection{Extinction Correction}
\label{subsection:EXT_COR}
To derive star formation rates and gas-phase metallicities, we also corrected for intrinsic extinction (on top of the foreground extinction correction performed initially). We used the Balmer decrement given by the flux ratio H$\alpha$/H$\beta$. This value has an intrinsic value of 2.86 (case B of \citealt{OsterbrockFerland2006}, assuming a temperature of T=10$^4$ K and electron density n$_\text{e}$=10$^2$ cm$^{-3}$). The colour-excess is computed via:

\begin{equation}
    E(B-V) = 1.97\,\text{log}\left(\frac{\text{H}\alpha/\text{H}\beta}{2.86}\right),
\end{equation}

\noindent{which further allows us to calculate the wavelength dependent extinction via:}

\begin{equation}
    A(\lambda)=k(\lambda) \times E(B-V),
    \label{eq:Ak}
\end{equation}
where $k(\lambda)$  is the value of the extinction curve at the given wavelength. As all galaxies in our sample host active star formation (Appendix~\ref{appendix-bpt}) we adopted the extinction curve calibration for starburst galaxies from \cite{Calzetti2000}. Using the extinction formula, each line was corrected via:

\begin{equation}
    F_{\text{int}}=F_{\text{obs}}10^{0.4A(\lambda)}
    \label{eq:flux}
\end{equation}


In the unbinned data at native MUSE resolution, we noticed that there are regions, predominantly within the outskirts of all the galaxies, that show a large scatter in the extinction from spaxel to spaxel. As this is caused by overall low signal in the Balmer emission lines, we set $A_\mathrm{V}$ = 0 for all bins with $E(B-V)$ > 0.8 or negative, similarly to \cite{curti_et_al_2020}. The rebinning of the data discussed in Sect. \ref{sect:rebinned} reduces the scatter. Additionally, we performed cuts on the spaxels where the H$\alpha$ S/N was lower than a threshold determined for each galaxy. The lowest threshold H$\alpha$ S/N adopted was $\approx$ 7.5.

\begin{figure*}[]
    \centering
    \includegraphics[width=0.9\textwidth]{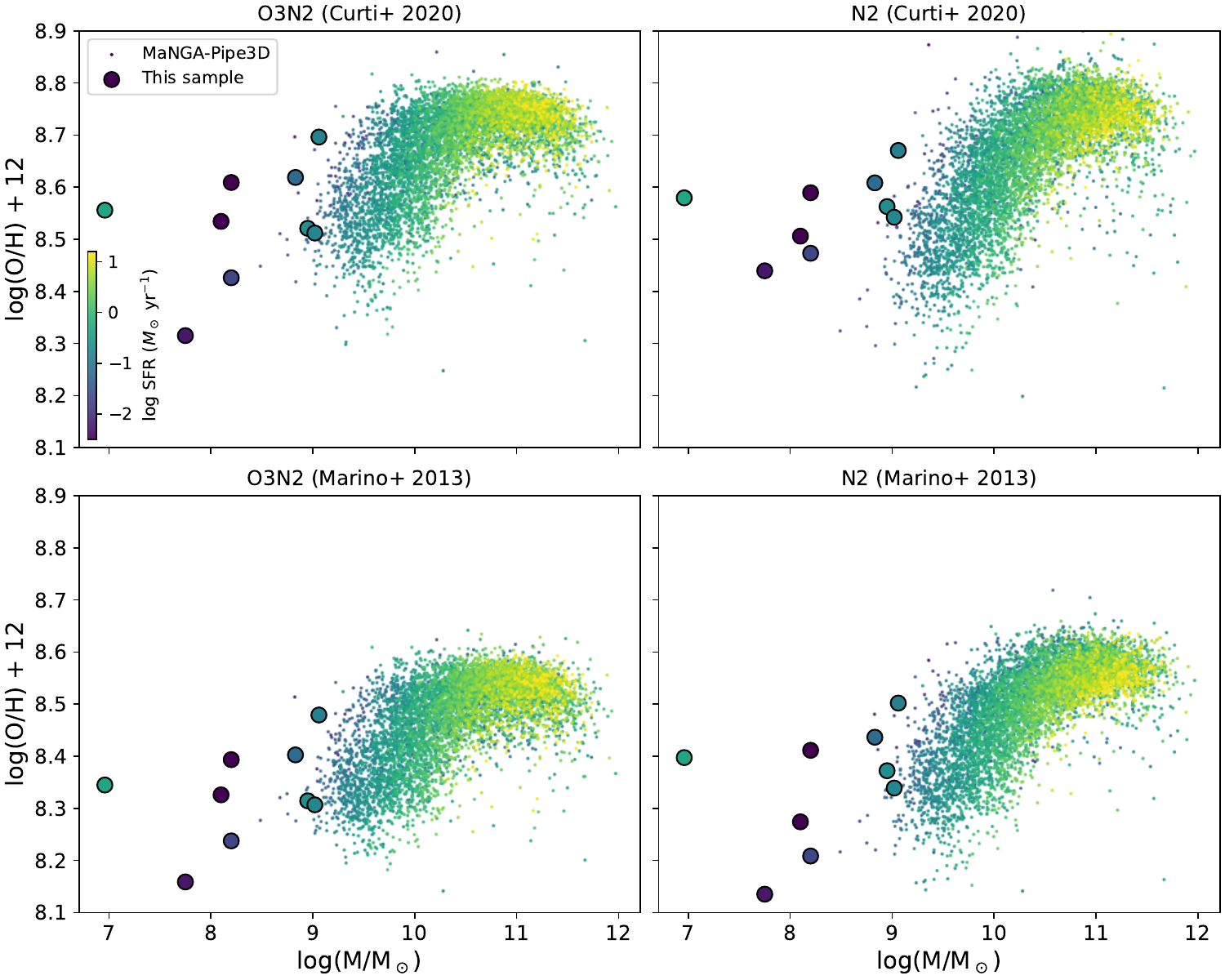}
    \centering
    \caption{Fundamental metallicity relation for our galaxy sample, in comparison with the MaNGA-Pipe3D galaxy catalogues for the SDSS seventeenth data-release \citep{mangapipe3d}, assuming four different metallicity calibrations -- O3N2 and N2 from \cite{curti_et_al_2020}, O3N2 and N2 from \cite{marino_et_al_2013}. The uncertainties in metallicity of our sample are smaller than the symbol sizes due to the high signal-to-noise of our integrated spectra.}
    \label{fig:FMR}
\end{figure*}

\subsection{Gas-phase Metallicity}
\label{subsection:metallicity}
To determine whether established metallicity calibrations can be safely used for our galaxies, we checked if the galaxies fall into the star forming regions in the Baldwin, Phillips $\&$ Terlevich (BPT; \citealt{BPT}) diagnostic diagram.  Figure~\ref{fig:BPT} in Appendix~\ref{appendix-bpt} shows that the galaxies are star forming. 
Consequently, we assume that all galaxies in our sample are star forming galaxies and this assumption will be used throughout the paper for all the subsequent analysis.

In order to compute the gas-phase metallicity, we used the O3N2 parameter, given by Eq.~\ref{eq:O3N2}, that involves the line-fluxes of H$\beta$($\lambda = 4861\AA$), OIII($\lambda = 5007\AA$), H$\alpha$($\lambda = 6562\AA$) and NII($\lambda = 6583\AA$): 

\begin{equation}
    \text{O3N2} = \text{log} \left( \frac{\text{[OIII]}_{5007}}{\text{H}\beta} \times \frac{\text{H}\alpha}{\text{[NII]}_{6583}}\right).
    \label{eq:O3N2}
\end{equation}
There has been extensive research on different calibration methods for O3N2 (e.g. \citealt{pettini_pagel_2004,maiolino_et_al_2008,marino_et_al_2013,brown_et_al_2016}). For the purpose of this paper however, we used the recent calibration proposed by \cite{curti_et_al_2017} and \cite{kumari_et_al_2019}, given by Eq.~\ref{eq:metallicity}.

\begin{equation}
    Z_g \equiv 12 + \text{log(O/H)} = 7.64 + 0.22 \times \sqrt{25.25-9.072 \times \text{O3N2}},
    \label{eq:metallicity}
\end{equation}
where $Z_g$ = 12 + log(O/H) represents the gas-phase metallicity, sometimes reported in the literature as oxygen abundance. For the rest of the paper this will be referred to as metallicity or simply $Z_{g}$. 

We noticed that different calibration methods for O3N2, as the ones specified before, give rise to the same overall metallicity trends, but with some offsets. As explained in \cite{curti_et_al_2017}, the O3N2 calibration method is well defined for 12 + log(O/H) values between 7.6 and 8.85, which is the case for the nine galaxies considered in this study. We also explored the S2, N2 and O32 calibrations, finding similar results and establishing that the O3N2 calibration is an appropriate option for our galaxy sample (see also Fig.~\ref{fig:FMR}, Sect.~\ref{overview-maps}).

\subsection{Star Formation Rate}
The star formation rate in each spaxel was estimated in the same way as in \cite{Fahrion2022}, using the relation from \cite{Hao2011}, relating the SFR to the intrinsic luminosity of the H$\alpha$ line:
\begin{equation}
    \text{SFR}(\text{M}_\odot\,\text{yr}^{-1}) = 10^{-41.257} L_\text{int}(\text{H}\alpha),
    \label{eq:SFR}
\end{equation}
with $L_\text{int}(\text{H}\alpha$) in erg s$^{-1}$. This relation assumes a \cite{kroupa} IMF. The luminosity was inferred from the extinction-corrected flux, as described in Sect.~\ref{subsection:EXT_COR}. By accounting for the pixel size of MUSE at the respective distance, we obtained the star formation rate densities in M$_\odot\,\text{yr}^{-1}$ kpc$^{-2}$. 
As our galaxy sample contains several irregular galaxies, we do not apply any inclination corrections as the inclination is challenging to measure in such galaxies. As such, there will be an intrinsic offset in the SFR results. Nonetheless, our sample was selected by having well visible nuclear star clusters in them and that points to the fact that these are not seen edge-on. Hence the corrections due to unknown inclination are likely to not to affect our main analysis significantly in the SFR results between different galaxies. Nonetheless, comparisons of regions within the same galaxy are not affected by this.

\section{Fundamental Metallicity Relation}
\label{section:FMR}

In this section, we investigate how our results compare with the FMR predictions, analysing what properties might result in deviations from the FMR.
\subsection{Overview of the Galaxy Sample}
\label{overview-maps}

We show our galaxies on the global FMR in Fig.~\ref{fig:FMR}, in comparison with the SDSS 17 data-release results, as presented in the MaNGA-Pipe3D catalogues \citep{mangapipe3d}. 
To obtain global gas-phase metallicities, we collapsed the MUSE cubes into a single spectrum for each galaxy, thereby simulating an unresolved measurement. This single spectrum was fitted with DAP and we used four different metallicity calibrations -- O3N2 and N2 from \cite{curti_et_al_2020} and \mbox{\cite{marino_et_al_2013}} for putting our galaxies in context. Whilst the overall characteristics are independent on the metallicity calibration used, we note that there is an offset in the metallicity values obtained from different relations. As explained in Sect.~\ref{subsection:metallicity}, for all the subsequent analysis of this study, we rely on the \cite{curti_et_al_2020} O3N2 prescription, as it is one of the most up-to-date and revised relation. Lastly, it can be seen that despite the small size of our galaxy sample, they make a significant contribution in the area of the less explored low mass and low SFR regime, extending the results from larger galaxies samples. We remind the reader however that the SFR results are not corrected for inclination, hence present an offset from the true values. We also note that, despite the low-mass regime explored in this study, the range of $\Sigma_{\mathrm{SFR}}$ and $\Sigma_{*}$ covered by our galaxies are comparable to surveys of disk galaxies (e.g. \citealt{sanchez2013}). We note however that because the surface mass density is projected, comparing the results with thinner structures such as disk galaxies is difficult. 



\begin{figure*}[]
    \centering
    \includegraphics[width=1\textwidth]{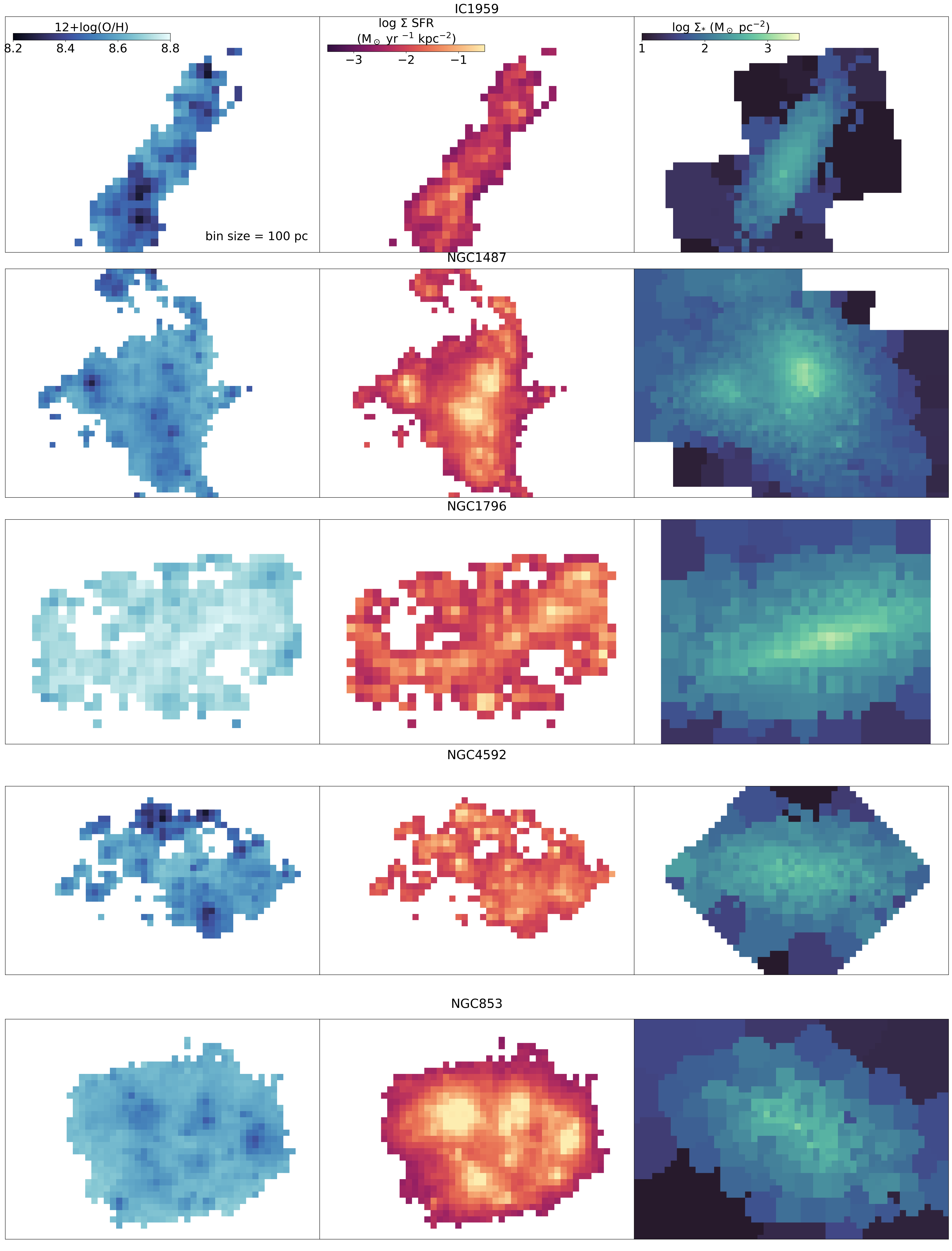}
    \caption{From left to right: gas-metallicity, star formation rate surface density and stellar mass surface density maps for five of the galaxies in our sample. The flux cuts derived on the binned H$\alpha$ data were enforced only on the metallicity and SFR maps.}
    \label{fig:amazing}
\end{figure*}

Furthermore, we show a qualitative representation of the resolved fundamental metallicity relation (rFMR) in our galaxy sample in Fig.~\ref{fig:amazing}. This figure presents the maps of the gas-phase metallicity, star formation surface density and stellar mass surface density for five out of nine galaxies in our sample. The five galaxies were chosen because they present enough bins for a qualitative rFMR analysis. The stellar mass surface density maps are shown without any flux cuts, while the other two incorporate the cuts described in Sect.~\ref{subsection:EXT_COR}. Despite the fact that we binned our data to a spatial resolution of 100 pc, Fig.~\ref{fig:amazing} shows a variety of diverse features both within individual galaxies as well as when comparing among galaxies. A general, qualitative $Z_{g}$-$\Sigma_{*}$ correlation  is present in the maps on a global scale. This correlation is not as obvious on small scales, and the maps illustrate the complex structures within galaxies. For example, in the case of NGC\,1487 there are regions in the central and northern parts where the two quantities are anti-correlated. We propose that the reasons behind this lies in the fact that NGC\,1487 is a merger, hence describing the overall behaviour of this galaxy is challenging.

It can also be seen that regions with low metallicity correspond to high star formation rate densities and vice-versa, in agreement with the FMR. However, this is not the case for NGC\,1796, where especially the central bar presents both a high metallicity and star formation rate density. NGC\,4592 shows regions in the centre where $Z_{g}$ and $\Sigma_{\mathrm{SFR}}$ appear to be correlated, but a clear anti-correlation is found in the off-plane star forming regions of this galaxy.

While the maps already provide insights into the complexity and diversity of the FMR on local scales, in the following subsections we aim to explore: i) how the $\Sigma_{\mathrm{SFR}}$-$\Sigma_{*}$ correlation behaves for the different galaxies in our sample and what contribution (if any) it has to the FMR; ii) the $Z_{g}$-$\Sigma_{*}$ and $Z_{g}$-$\Sigma_{\mathrm{SFR}}$ dependencies in our galaxy sample and how they are affected by the galaxies' properties and iii) if there is a spatial dependence of the galaxies' properties in the cases where enough points are available and hence if it is rigorous to consider a merger galaxy such as NGC\,1487 in such a study.

\subsection {Star Formation Rate Surface Density - Stellar Mass Surface Density Dependence}
\label{sect:SFR-sigma}

\begin{figure*}
    \centering
    \includegraphics[width=0.95\textwidth]{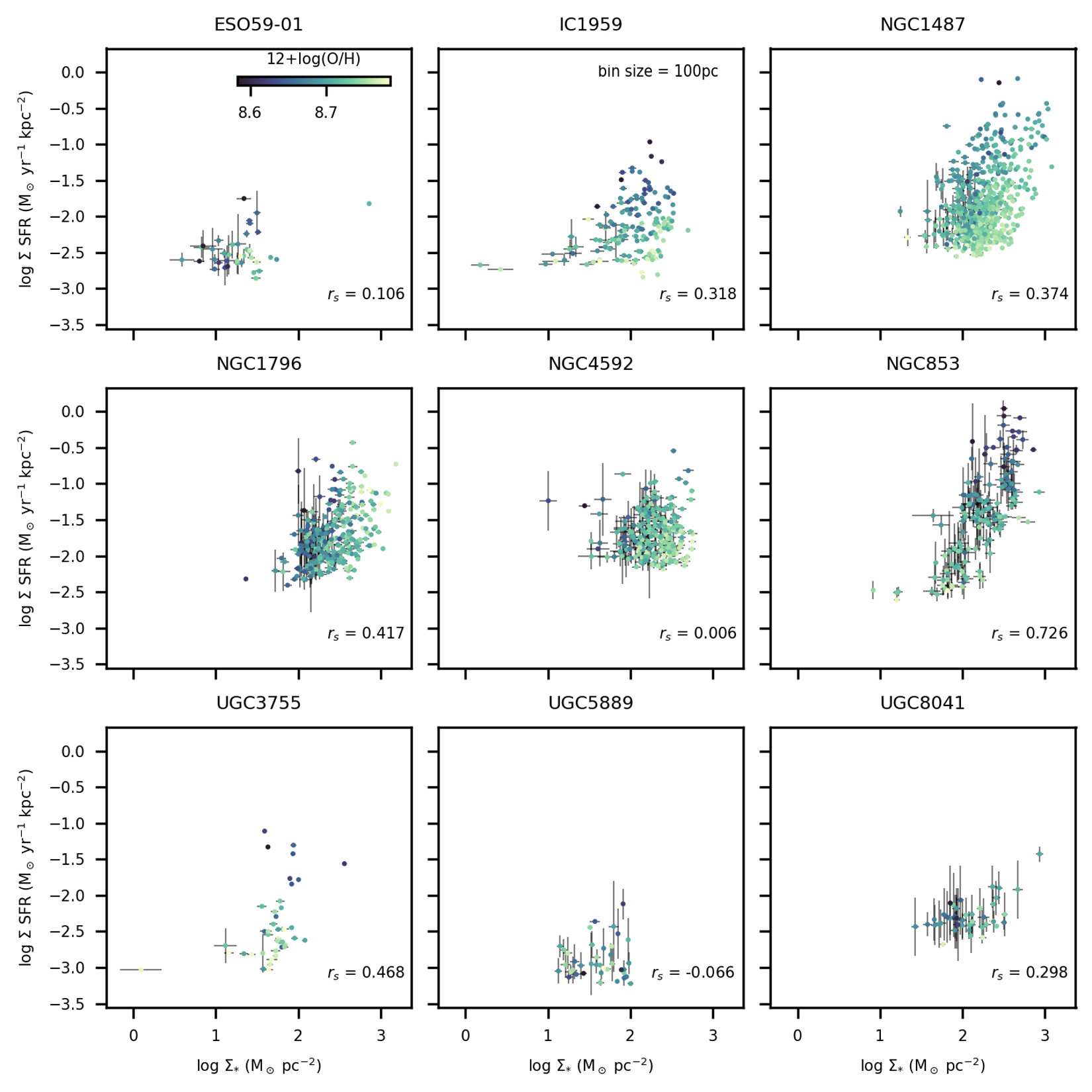}
    \caption{Star formation rate surface density dependence on the stellar mass surface density for the galaxy sample, binned according to the stellar mass density bins. The points are colour-coded by the gas-phase metallicity. Errorbars show the uncertainty of the surface mass density from the DAP fits in each bin as well as the standard deviation of star formation rate density values in each bin.} The Spearman's correlation coefficient, $r_s$ is included in all the plots.
    \label{fig:SFR-sigma_9}
\end{figure*}

In this section, we focus on the $\Sigma_{\mathrm{SFR}}$-$\Sigma_{*}$ dependence for each galaxy, shown in Fig.~\ref{fig:SFR-sigma_9}. As explained in Sect.~\ref{sect:rebinned}, for this analysis, both $\Sigma_{\mathrm{SFR}}$ and $\Sigma_{*}$ are computed for binned data, with a bin size of 100 pc. The Spearman's correlation coefficients ($r_s$) are shown in all the plots.   

Figure~\ref{fig:SFR-sigma_9} shows an overall correlation between $\Sigma_{\mathrm{SFR}}$-$\Sigma_{*}$. Observational limitations do not allow for decisive conclusions in the cases of ESO\,59-01, UGC\,5889 and UGC\,8041. Moreover, the different behaviours within NGC\,4592, observed qualitatively in Fig.~\ref{fig:amazing}, lead to significant scatter and no statistical evidence for an overall correlation in this galaxy. To relate the $\Sigma_{\mathrm{SFR}}$-$\Sigma_{*}$ dependence with the FMR, all the points are colour-coded by $Z_{g}$ (also resulted from binned data, with a bin size of 100 pc). It can be seen that overall the high $\Sigma_{*}$ points correspond to high metallicities, but there is significant scatter (e.g. IC\,1959). The SFR-$Z_{g}$ trend is present only in some of the galaxies, as discussed in more detail in the next sections. 

We also note that the small range of the parameters, in particular $\Sigma_{*}$ represents a caveat of this study, as it does not allow us to make fair comparisons with the global FMR where these ranges are considerably larger. This can cause galaxies with low number of points and small $\Sigma_{*}$ range (e.g. UGC 5889) to not show a clear correlation.


\subsection {Metallicity - Stellar Mass Surface Density Dependence}
\label{sect:Z-sigma}

\begin{figure*}
    \centering
    \includegraphics[width=0.95\textwidth]{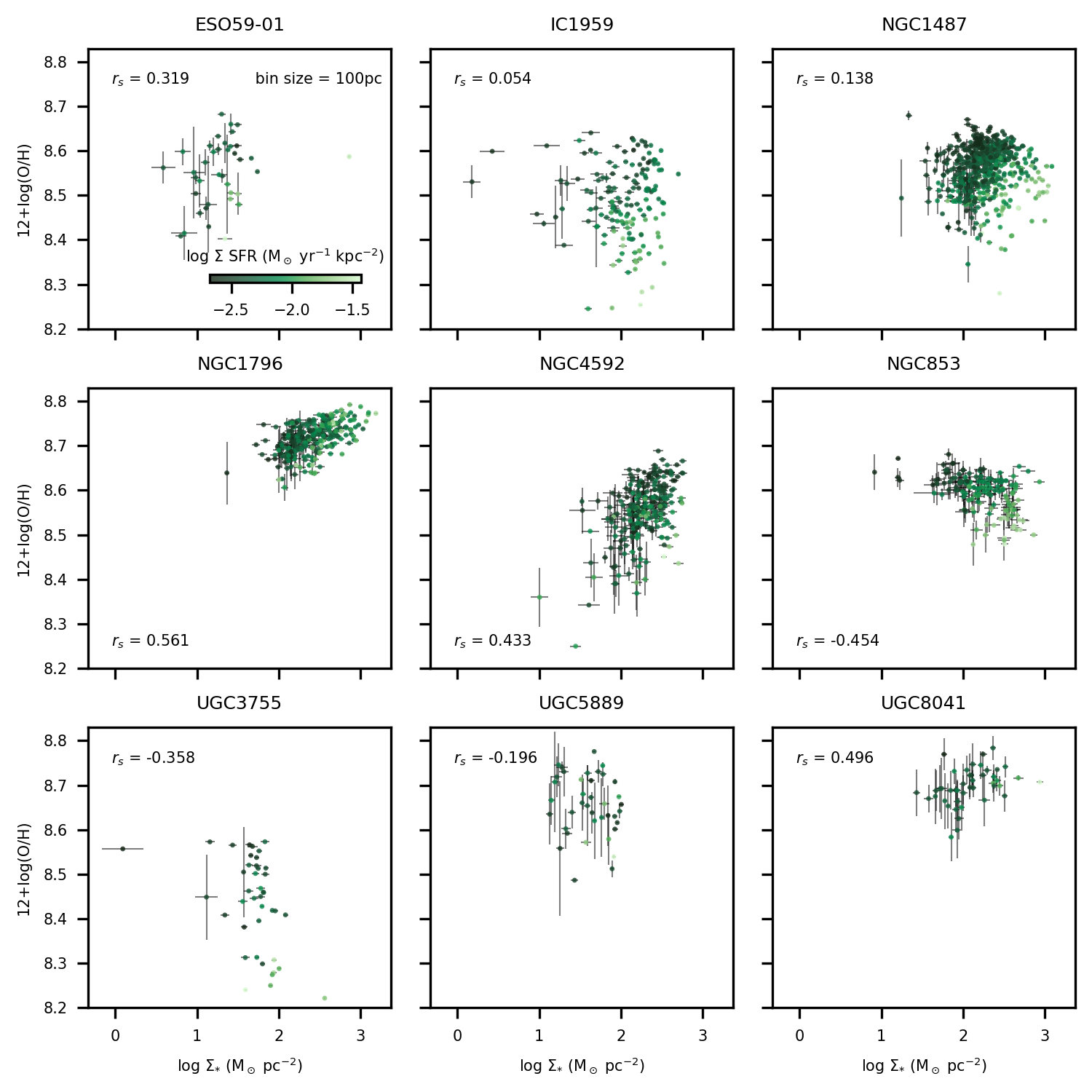}
    \caption{Gas-phase metallicity dependence on the stellar mass surface density for the galaxy sample, binned according to the stellar mass surface density bins. The points are colour-coded by the star formation rate surface density. Errorbars show the uncertainty of the surface mass density from the DAP fits in each bin as well as the standard deviation of metallicity in each bin.} The Spearman's correlation coefficient, $r_s$ is included in all the plots.
    \label{fig:met-sigma_9}
\end{figure*}

In this section, we focus on the $Z_{g}$-$\Sigma_{*}$ dependence for each of the galaxies in our sample, shown in Fig.~\ref{fig:met-sigma_9}. Both the metallicity and $\Sigma_{*}$ were binned, with a bin size of 100 pc and all the points are colour-coded by $\Sigma_{\mathrm{SFR}}$ in order to provide a better visual perspective on the comparison with the FMR. 

Similar to before, ESO\,59-01, UGC\,5889, UGC\,8041 present a lower number of points due to data limitations. It can be seen that the trends are not the same for all galaxies. NGC\,1796, NGC\,4592, UGC\,8041 appear to have the tightest correlations in the sample. For the lowest mass galaxies (NGC\,853 and UGC\,3755), the points present significant scatter, but also a different trend with regions that show a $Z_{g}$-$\Sigma_{*}$ anti-correlation, which leads to overall negative correlation coefficients. As explained in Sect.~\ref{sect:SFR-sigma}, the lack of correlation can also be caused by the small range of $\Sigma_{*}$ covered in these two galaxies. NGC\,1487 introduces the largest scatter of all the galaxies in the sample. Given that this galaxy is undergoing a merger, we propose that the characteristics of this galaxy might complicate the analysis in this case.

\subsection {Metallicity - Star Formation Rate Surface Density Dependence}

\begin{figure*}
    \centering
    \includegraphics[width=0.95\textwidth]{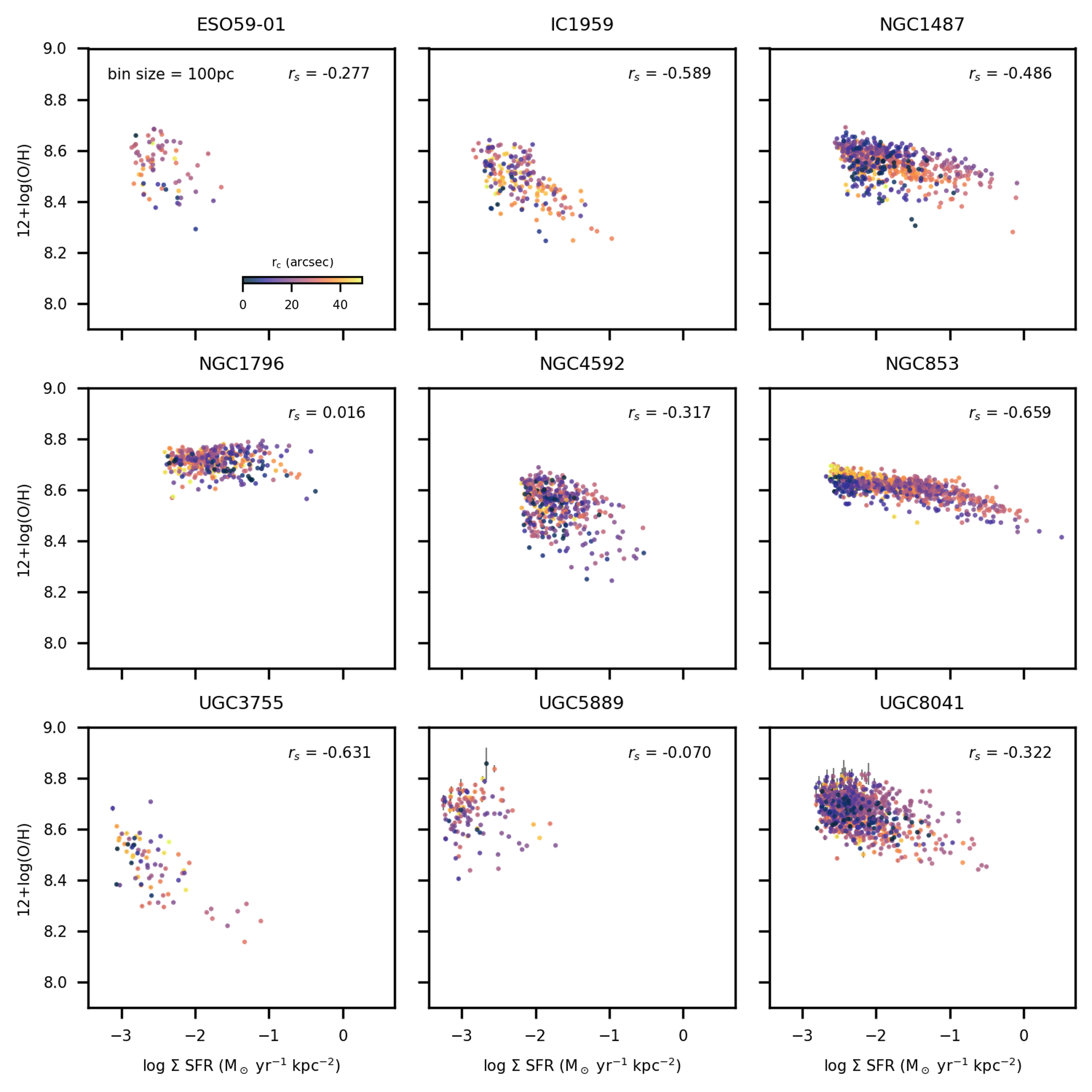}
    \caption{Gas-phase metallicity dependence on the star formation rate surface density for the galaxy sample. The plots include the binned data (bin size = 100 pc) obtained after the corrections, colour-coded by the distance from the galaxy centre r$_\mathrm{c}$. The Spearman's correlation coefficients $r_s$ are shown in all of the plots.}
    \label{fig:met-SFR_9_100}
\end{figure*}

In this section, we focus on the $Z_{g}$-$\Sigma_{\mathrm{SFR}}$ dependence. We show the binned data, with a bin size of 100 pc (Fig.~\ref{fig:met-SFR_9_100}). 
All of the points were colour-coded by the distance from the centres of the galaxies. This enabled us to investigate closer how different regions within the same galaxies behave. 
The coordinates of the centres of the galaxies are the nuclear star clusters and were obtained from \cite{Fahrion2022}. 

Figure~\ref{fig:met-SFR_9_100} shows that the galaxies in our sample present an overall anti-correlation between metallicity and SFR. The most massive galaxy in our sample (NGC 1796) shows a (small) positive correlation coefficient, which can be caused by the high mass of this galaxy. Additionally, UGC 5889 does not show clear signs of an anti-correlation, most likely due to the low number of points and observational limitations involved in this galaxy's properties. Nonetheless, in all cases the colour-coding of the points suggests that individual regions adopt behaviours dependent on the spatial location within the galaxies (most visible for the merger -- NGC 1487, but also for NGC 853). Due to the low spatial resolution of the data, after rebinning, these individual behaviours cannot be studied in great detail, but this shows the significant influence individual regions have on the averaged relations.

\section{Discussion and comparison with the literature}
\label{section:conclusions}
In all the dependencies analysed in Sect.~\ref{section:FMR}, we found significant scatter and in the cases of ESO\,59-01, UGC\,5889, UGC\,8041 a low number of points, due to the low emission line fluxes. For NGC\,1487 the scatter is also significant, due to its distorted morphology, stemming from its ongoing merger. However, we note that this scatter likely holds physical meaning as different regions within individual galaxies show different behaviours which leads to this scatter when considering the full galaxy.


The more debated quantity -- the SFR -- proved to play a non-negligible role in our analysis, however this feature varied significantly from galaxy to galaxy (e.g. Fig.~\ref{fig:met-SFR_9_100}). 
Overall, we found that the SFR is anti-correlated with the metallicity, this anti-correlation being generally stronger for the lower-mass galaxies in our sample, in agreement with the FMR. However, the galaxies show confined regions of star formation where there is a clear anti-correlation, each having its own slope. 
For the most massive galaxies in our sample ($\sim$10$^9$ M$_\sun$ galaxies), we found a weak anti-correlation, this effect being most evident in the case of NGC\,1796. 
Figure 34 from \cite{review_paper} shows that the overall SFR-$Z_{g}$ anti-correlation flattens out for high stellar mass surface densities, in qualitative agreement with our findings (Fig.~\ref{fig:met-sigma_9}). \cite{sanchez_menguiano_et_al_2019} also reported that the SFR-metallicity anti-correlation behaves differently in different galaxies. Studying a total of $\sim$ 700 star-forming nearby spiral galaxies from the MaNGA survey, they found that 60$\%$ of the galaxies present an anti-correlation, 19$\%$ show no correlation and the rest of 21$\%$ present a positive correlation. They proposed that the SFR-metallicity anti-correlation comes from low-mass metal-poor galaxies, while higher masses and more metal-rich systems exhibit a reverse trend. They concluded that the driver of this finding is the average gas-phase metallicity of the galaxy, in the sense that metal-poor systems show stronger anti-correlations. \cite{Laura} reported the same trend via EAGLE simulations using a sample of 107 simulated disk galaxies. 

It can be seen that the highest metallicity galaxy (NGC 1796) shows indeed the weakest anti-correlation. Also, for specific regions with the galaxies, the SFR-Zg profile becomes flatter at higher metallicities. This effect is most visible in the case of NGC 1487 (Fig.~\ref{fig:met-SFR_9_100}). Moreover, NGC\,853, the least massive galaxy in our sample shows a considerably different behaviour than the other galaxy of similar mass in our sample (UGC\,3755). Additionally, as previously seen, this galaxy deviates significantly from the global FMR predictions (Fig.~\ref{fig:FMR}). These features can indeed be explained by NGC\,853's untypical gas-phase metallicity for its assumed mass. However, a larger sample studied at these high spatial resolutions would be needed to confirm this trend.

In agreement with \cite{sanchez_menguiano_et_al_2019} and \cite{Laura}, we note that our results support a scenario in which external metal-poor gas accretion drives the star formation in metal-poor galaxies, while in metal-rich systems the gas comes from previous star forming episodes, hence being more metal-enriched.

Moreover, we investigated the connection between SFR and stellar mass surface density (Fig.~\ref{fig:SFR-sigma_9}), finding a considerable overall correlation, in agreement with the well-known SFMS. Given the MZR, it is not unreasonable to observe gradually weaker SFR-$Z_{g}$ anti-correlations for increasingly higher stellar masses. Due to the limited number of galaxies in our sample, it is not possible at this stage to confirm if there is a stellar mass surface density, or rather an average metallicity threshold where the SFR-$Z_{g}$ anti-correlation breaks down, but there is definitely scope to investigate this via larger surveys with a comparable resolution. 

We also showed that for our unbinned data, the SFR-$Z_{g}$ dependence behaves differently for different regions within the galaxies, this effect being most evident in the cases of NGC\,1487 and NGC\,1796 (Fig.~\ref{fig:met-SFR_9_100}). We discovered that individual regions play an important role in shaping the overall SFR-$Z_{g}$ dependence. Hence, the high spatial resolution employed in this study proves the importance of investigating spatially-resolved galaxies, as crucial information can hide within the small-scale spatial characteristics of these galaxies. 

Despite the consistent effort and rapid technological advancement (both for observations and simulations), some open issues with regards to the FMR still remain. For instance, having to rely on strong-line diagnostics represents one of the main limitations of such studies. \cite{curti_et_al_2020} implemented and compared different electron temperatures calibrations in the context of the FMR, hence opening new avenues for more accurate estimations, but there is currently a lack of universally accepted calibration methods. Moreover, \cite{kewley_2007} and \mbox{\cite{curti_et_al_2020}} showed that the choice of metallicity calibration is crucial, as it strongly influences the parameters of the FMR. We also caution that \cite{teklu_et_al_2020} showed, using data from the MaNGA survey, that the O3N2 calibration leads to results in agreement with the FMR, while the photoionization model-dependent N2O2 and N2S2 calibrations introduce unexpected results in comparison to the FMR. For our study, we chose the most updated versions of the O3N2 calibration presented in \cite{curti_et_al_2017} and \cite{kumari_et_al_2019}. 

Moreover, \cite{curti_et_al_2022} showed via the recent JWST data that current metallicity calibrations in the literature do not universally apply to all the galaxies and appear to be considerably influenced by redshift also. There is scope for further investigation in this direction. Particularly, the JWST and Extremely Large Telescope (ELT) are expected to provide an unequalled resolution, enabling an in-depth study of small-scale spatially-resolved properties. However, these two telescopes operate mainly in the near- and mid-infrared, which restricts their use for similar studies in the Local Universe, because the metallicity calibrations are based on optical lines. On the other hand, they can apply the calibrations to high-redshift galaxies.

MUSE studies can be pushed even further (e.g. MUSE Atlas of Disks - \citealt{erroz-ferrer_et_al_2019}, PHANGS-MUSE - \citealt{Emsellem_et_al_2022}) by increasing the number of galaxies studied and extending the analysis to lower masses. Additionally, both large-volume simulations (e.g. SIMBA -- \citealt{simba}, EAGLE -- \citealt{crain_et_al_2015,schaye_et_al_2015}, ILLUSTRIS TNG -- \citealt{pillepich_et_al_2018},
HORIZON-AGN -- \citealt{dubois_et_al_2014}) and zoom-in simulations (e.g. VELA -- \citealt{ceverino_et_al_2014}, \mbox{\citealt{zolotov_et_al_2015}},
NIHAO -- \citealt{wang_et_al_2015}, APOSTLE
-- \citealt{sawala_et_al_2016} or LATTE/FIRE -- \citealt{wetzel_et_al_2016}) have registered an accelerated progress over the last couple of years (for a detailed review, see \citealt{vogelsberger_et_al_2020}) and are expected to build new grounds on the topic of FMR, especially regarding the factors influencing it.

\section{Conclusions}
\label{conclusions_2.0}
This paper presents a spatially resolved analysis of star formation and enrichment processes in a sample of nine dwarf galaxies. This study opens new avenues on the topic of star formation at low metallicity in dwarf galaxies due to the exceptional sensitivity of MUSE. Our conclusions are summarised in the following:

\begin{itemize}

   \item{We find significant scatter within the individual galaxies and in some cases the small $\Sigma_{*}$ ranges cause lack of correlations. Hence a fair comparison with the global FMR cannot be achieved by such a study, but the importance of individual galaxies and individual regions within galaxies is evident in our analysis.}
    \item The SFR-$Z_{g}$ dependence adopts several different slopes within individual galaxies, this effect being more evident in the cases of NGC\,1487 and NGC\,853. Hence studying spatially resolved quantities is crucial in this context and anticipates high uncertainties when averaging them on large scales, with a much larger scatter than the error bars in SFR and $Z_{g}$ (Fig.~\ref{fig:met-SFR_9_100}).
    \item The SFR-$Z_{g}$ dependence is also influenced by the total stellar mass, in agreement with the FMR, and by the average metallicity. As such, overall the lower mass and/or the lower metallicity galaxies present a tighter SFR-$Z_{g}$ anti-correlation.
    \item The merger state of NGC\,1487 should not be neglected, as demonstrated by the fact that this galaxy deviates significantly from the expected correlation between surface mass density and metallicity.
\end{itemize}

MUSE's high sensitivity, together with the low masses of our sample, allows us to identify signatures of the fundamental metallicity relation in the dwarf galaxy regime.  While the typical mass and SFR densities and are comparable to surveys of star forming disk galaxies, the different structural properties and bursty star formation histories for low mass dwarf galaxies reveal significant diversity in how SFR, metallicity and mass density are locally connected. Our findings agree with the hypothesis that the gas fuelling star formation is metal-poor in low mass galaxies, while in massive metal-rich systems this gas is recycled continuously and hence more metal-enriched. All in all, it is clear that we have entered an exciting era to study the chemical enrichment processes within galaxies with the design of new telescopes and the development of simulations. In particular, the galaxies used in this study could represent local analogues of resolved high-z galaxies form state-of-the-art JWST studies. 

\begin{acknowledgements}
      The authors would like to thank the referee for the detailed and constructive feedback, as well as Francesco Belfiore and Nimisha Kumari for helpful discussions. TEB acknowledges support through the Leiden/ESA Astrophysics Program for Summer Students (LEAPS) 2022. KF and FM acknowledge support through the ESA Research Fellowship programme. SRON is supported financially by NWO. OM is grateful to the Swiss National Science Foundation for financial support under the grant number PZ00P2\_202104. GvdV acknowledges funding from the European Research Council (ERC) under the European Union's Horizon 2020 research and innovation programme under grant agreement No 724857 (Consolidator Grant ArcheoDyn) This project makes use of the MaNGA-Pipe3D data products described in \cite{Sanchez2016, mangapipe3d}. We thank the IA-UNAM MaNGA team for creating this catalogue, and the Conacyt Project CB-285080 for supporting them.
\end{acknowledgements}


\bibliographystyle{aa} 
\bibliography{References} 
\newpage
\appendix

\section{BPT Diagrams}
\label{appendix-bpt}
\begin{figure*}[h!]
    \centering
    \includegraphics[width=0.95\textwidth]{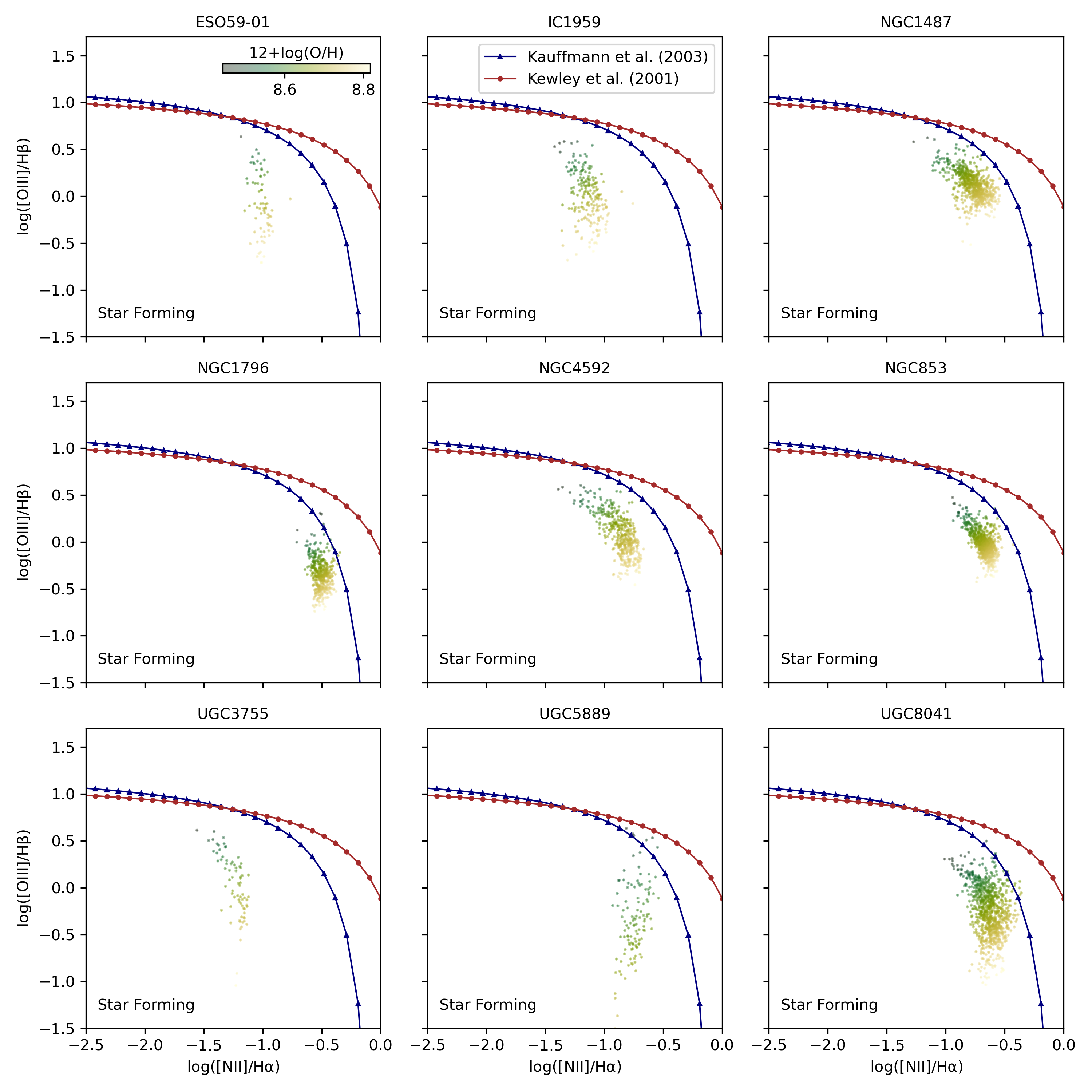}
    \caption{Baldwin, Phillips $\&$ Terlevich (BPT) diagnostic diagrams for the galaxy sample, with two literature models -- \cite{kewley_et_al_2001} (red curve) and \cite{kauffmann_et_al_2003} (blue curve). The star forming region lies below the two model curves. The points are obtained from binned data, with a bin size of 100 pc and colour-coded by the gas-phase metallicity -- 12 + log(O/H).}
    \label{fig:BPT}
\end{figure*}

Figure~\ref{fig:BPT} shows the BPT analysis for the nine galaxies in our sample. The plots show that the galaxies in our sample are star forming. 

\end{document}